\documentclass[aps,twocolumn,showpacs,amsmath,amssymb,pra,superscriptaddress,floatfix,longbibliography]{revtex4-1}

\usepackage{graphicx}    
\usepackage{hyperref} 
\usepackage{xcolor} 
\usepackage{physics} 
\usepackage{subcaption} 
\usepackage{ragged2e}
\usepackage[justification=justified, singlelinecheck=false]{caption}
\usepackage{bm}
\usepackage{booktabs}
\usepackage{soul}

\usepackage{xargs}
\usepackage{dcolumn}
\usepackage[colorinlistoftodos,prependcaption,textsize=tiny]{todonotes}
\newcommandx{\tohfa}[2][1=]{\todo[linecolor=OrangeRed,backgroundcolor=OrangeRed!25,bordercolor=red,#1]{#2}}
\newcommandx{\fr}[2][1=]{\todo[linecolor=Blue,backgroundcolor=Blue!25,bordercolor=blue,#1]{#2}}

\newcommand{\imu}{\text{\rm i}}

\newcommand{\spindown}[1]{\hat\sigma_{#1}}
\newcommand{\spinup}[1]{\spindown{#1}^\dagger}
\newcommand{\exc}[1]{\hat e_{#1}}

\newcommand{\fieldmode}[2]{\hat{#1}_{#2}}

\newcommand{\wgmode}[1]{\fieldmode{a}{#1}}

\newcommand{\wgbmode}[1]{\fieldmode{b}{#1}}

\newcommand{\betab}{\beta^-}
\newcommand{\betaf}{\beta^+}

\newcommand{\betaUni}{\beta}

\newcommand{\density}{\hat {\rho}}

\newcommand{\ddt}{\frac{d}{dt}}
\newcommand{\ha}{\text{h.c.}}

\newcommand{\nth}[1]{#1^\text{th}}
\newcommand{\expect}[1]{\langle #1 \rangle}
\newcommand{\Eq}[1]{Eq.~(#1)}

\newcommand{\allexcited}{\ket{ee\dots e}}
\newcommand{\B}{\mathcal{B}}

\begin{document}
\title {Photon statistics in waveguide QED: II Exact solutions in a thermodynamic limit}

\author{M. Eltohfa}
 \email{meltohfa@purdue.edu}
 \affiliation{
 Department of Physics and Astronomy, Purdue University, West Lafayette, Indiana 47906 USA
}
\author{F. Robicheaux}%
 \email{robichf@purdue.edu}
\affiliation{
 Department of Physics and Astronomy, Purdue University, West Lafayette, Indiana 47906 USA
}
\affiliation{Purdue Quantum Science and Engineering Institute, Purdue
University, West Lafayette, Indiana 47907, USA}
\date{\today}


\begin{abstract}
    Waveguide quantum electrodynamics (WQED) offers a powerful framework for controlling light–matter interactions and realizing collective phenomena such as super- and subradiance. In general waveguide settings, the quantum dynamics spans the full Hilbert space, rendering exact theoretical treatments exponentially difficult and currently out of reach, and only a few models have exact, analytical solutions. Motivated by recent experiments, we treat the thermodynamic limit of the number of atoms, $N \rightarrow \infty$, while the homogeneous atom–waveguide coupling $\beta \rightarrow 0$ keeping the optical depth $4N\beta$ fixed. In this limit, a second order mean field method is exact, giving analytical solutions for the statistics of the photons emitted in the waveguide both for chiral and symmetric configurations starting from full inversion. As $N \rightarrow \infty$, the emission in freespace approaches that of an independent ensemble. However, until a special time, $\approx 1.59 \times$ the lifetime of a single-atom, we show an exponentially enhanced superradiance in the waveguide as the optical depth increases. After the special time, the emission into the waveguide exhibits subradiance. We also show that the initial shot-to-shot fluctuations in the rate of emission into the waveguide diminish in a chiral system and vanish in a symmetric system as $N$ approaches $\infty$. Additionally, the equal-time second-order correlation becomes trivial, showing that finite-size effects are essential to observe the emergence of second-order coherence. Finally, going beyond the thermodynamic limit requires higher order mean field methods. Our results illustrate finite- and infinite-body collective effects in symmetric and symmetry-lacking systems.
\end{abstract}

\maketitle

\section{Introduction}\label{sec:intro}

Waveguide quantum electrodynamics (WQED) offers a suitable environment to achieve interesting collective atom-light phenomena such as super- and subradiance \cite{dicke1954coherence,gross1982superradiance,gross1976observation,scully2009collective,pellegrino2014observation,prasad2000polarium,devoe1996observation,RevModPhys.95.015002}. These have been demonstrated in various theoretical \cite{asenjo2017exponential,TWA2024cascaded,anaPhysRevLett.131.033605,Retardation2025effects,MFchiral2023higherorder} and experimental \cite{setupPhysRevLett.104.203603,PRLLiedl2023collective,PRXQuantum.4.030304,bach2024emergence} studies. In this work, we focus on the statistics of the photons emitted from an inverted ensemble of waveguide-coupled atoms. This includes the photon flux and two-time photon-photon correlation functions, which have been experimentally demonstrated \cite{bach2024emergence}. Unlike in the Dicke limit, where there is a permutational symmetry between all atoms, a general WQED setup lacks any symmetry \cite{MFchiral2023higherorder} resulting in the system exploring all the available states of the Hilbert space. This was demonstrated in the single-excitation manifold \cite{oscillationPhysRevLett.128.203601}. In the many-excitation scenario considered in this work, this implies the need to simulate an exponentially growing Hilbert space to obtain exact solutions, which is a computational challenge.

In a companion paper \cite{tohfachiral}, we tackled this challenge by developing an efficient higher-order mean field (MF) approach that is able to model the photon statistics in a recent experiment in Ref.~\cite{bach2024emergence}. This experiment is done for a chiral system, where many atoms ($N \sim 1000$) mainly couple to the forward propagating mode of the waveguide with a small coupling ($\beta \sim 0.01$). Moreover, we developed an analytical approach that is scalable for a general atom number, $N$, and a homogeneous coupling strength starting from an inverted state. The accessible regime of the experimental parameters in Ref.~\cite{bach2024emergence} and the scalability of our analytical approach motivate us to consider a thermodynamic limit where $N \rightarrow \infty$ with the optical depth, $OD = 4 N \beta$, fixed.

In this paper, we study such thermodynamic limit for a chiral and a permutationally symmetric systems. We find that the complicated \emph{finite} $N$ analytical solution developed for a chiral system in the companion paper \cite{tohfachiral} collapses to a simple form for an \emph{infinite} $N$. Moreover, we find that a second-order mean field (MF2) is exact in this limit. This lends a way for a simple interpretation of the symmetry-lacking chiral system as a continuous medium, whose observables have a closed form (or easy to compute) analytical solutions.

Predictions can be extracted from the analytical solutions. 1) The power emitted into the waveguide grows exponentially as $N \beta$ is increased. 2) For a \emph{chiral} system, the second order coherence $g^{(2)}(0, t)$ \cite{bach2024emergence} stays nearly flat for all times around the superradiance peak. This shows that the shot-to-shot fluctuations in the photon rate demonstrated in the experiment \cite{bach2024emergence} diminish in the thermodynamic limit. However, the shot-to-shot fluctuations increase with $N \beta$. This is in contrast with a \emph{permutationally symmetric} system where $g^{(2)}(0, t) = 2$ for all $t$ for all $N \beta$. 3) the \emph{same-time} second order coherence $g^{(2)}(t, t) = 2$ for both a chiral and a permutationally symmetric systems.

We note that another limit $N \rightarrow \infty$ at \emph{fixed} non-zero $\beta$ was derived for a symmetric system in Ref.~\cite{NlimitPhysRevA.106.013716}. In this limit, \emph{almost all} the energy stored in the atoms is radiated into the waveguide despite the presence of freespace decay, and the dynamics approaches that of pure Dicke superradiance ($\beta$=1). The limit in this work is different and entails that $\beta \rightarrow 0$, and is also derived for a symmetry-lacking chiral system. In such limit, the fraction of the energy emitted in the waveguide goes to $0$. Practically for a finite $N$, the $\beta \rightarrow 0$ limit is a good approximation for the radiation dynamics as long as the fraction of energy radiated into the waveguide is $\lesssim 1\%$. This was numerically checked for a symmetric system for a wide range of parameters.

This paper is organized as follows. In Sec.~\ref{sec:methods}, we define the system under investigation and the associated MF equations. In Sec.~\ref{sec:analytical_methods}, we examine the thermodynamic limit of the analytical solution for a chiral system starting from full inversion. In Sec.~\ref{sec:continuum_approximation}, we present the continuum approximation of the second order mean field (MF2) method for a chiral system and show its exactness by comparing it to the solution from Sec.~\ref{sec:analytical_methods}. In Sec.~\ref{sec:continuum_approximation_sym}, we apply the thermodynamics limit of the MF2 for a permutationally symmetric system which results in simpler dynamics than a chiral system. In Sec.~\ref{sec:approaching_therm_limit}, we explore the approach to the thermodynamic limit for the photon rates and second-order coherence. In Sec.~\ref{sec:results_thermodynamic_limit}, we explore the aforementioned predictions regarding the thermodynamic limit. In Sec.~\ref{sec:corr_exc}, we examine the buildup of correlations and the enhanced excitation decay for the chiral system. In the appendix, we show a mathematical proof of the exactness of MF2 for a symmetric system, and investigate the required order of mean field to explore corrections beyond the thermodynamic limit.



\section{Methods}
\label{sec:methods}

\begin{figure}[htbp]
    \centering
    \includegraphics[width=0.45\textwidth]{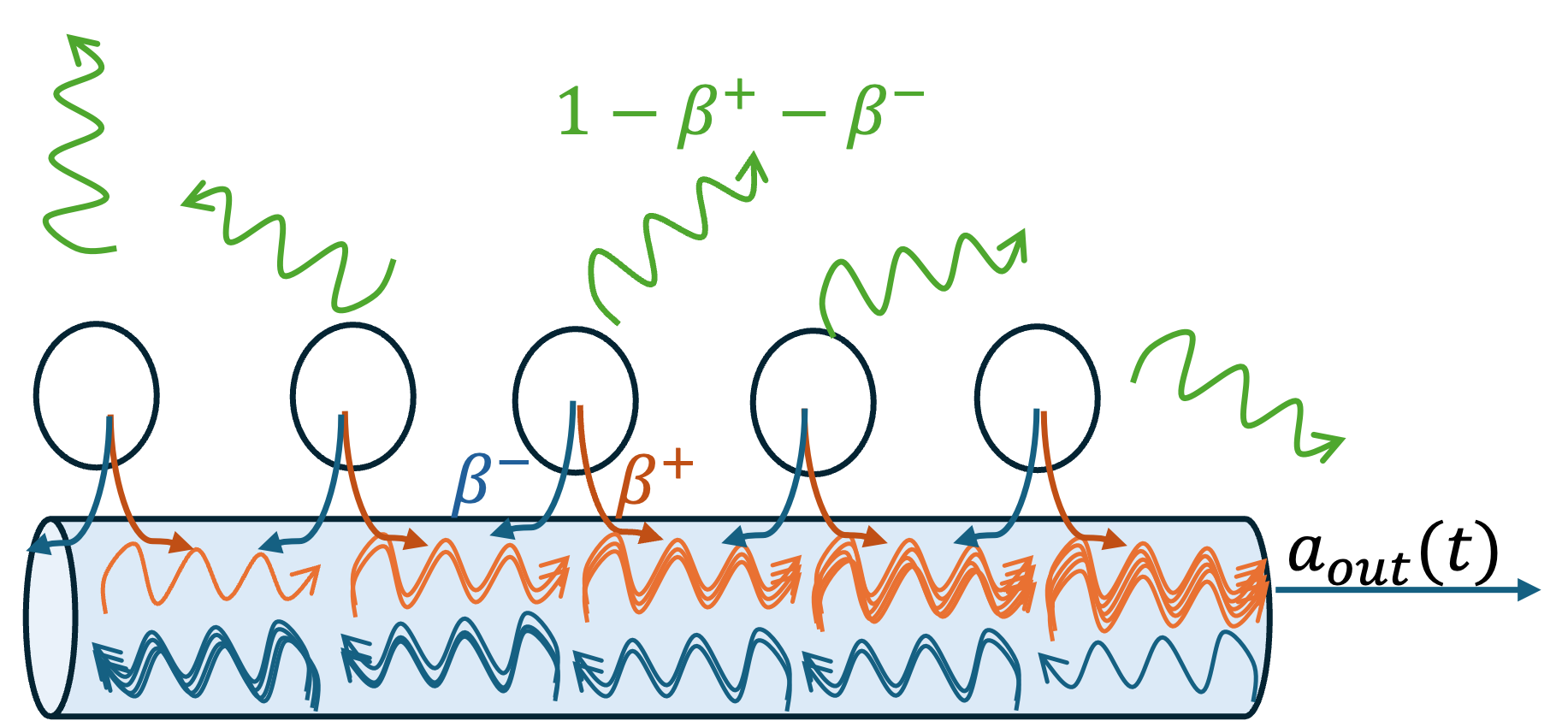}
    \caption{\justifying
        Schematic of the setup. $N$ two-level atoms are coupled to a waveguide.
        The coupling constant to the right waveguided mode is $\beta^+$, to the left waveguided mode is $\beta^-$, and to freespace modes is $1 - \beta^+ - \beta^-$. This paper investigates two extreme cases: 1) a unidirectional waveguide ($\betab = 0$, $\betaf = \beta$) and 2) a bidirectional and permutationally symmetric case where $\betaf = \betab = \betaUni/2$ and the atoms are placed in the mirror configuration with $\pi$ or $2 \pi$ phase between neighboring atoms.
        }
    \label{fig:intro_figure}
\end{figure}

The system under consideration is described in detail in Refs.~\cite{PRLLiedl2023collective,PRXsuperradiant2024bursts,TWA2024cascaded} and is shown in Fig.~\ref{fig:intro_figure}. It consists of a waveguide with $N$ two-level atoms trapped in its vicinity. The excited state is denoted as $\ket{e}$, while the ground state is denoted as $\ket{g}$. We denote the ordered atom positions along the waveguide as $z_i$ for $ 1 \leq i \leq N$ such that $z_i < z_j$ if $i < j$. A single atom radiates spontaneously with rate $\Gamma_0$ which branches into the waveguide's forward and backward directions and into freespace modes with probabilities $\betaf$, $\betab$, and $1 - \betaf - \betab$ respectively. In this paper, we treat two cases: 1) a unidirectional (also referred to as chiral) waveguide ($\betab = 0$), and denote $\betaf \equiv \betaUni$ and 2) a bidirectional and permutationally symmetric case, $\betaf = \betab \equiv \betaUni/2$. Also, we assume that every atom has the same coupling.

As is commonly considered, we work in the Markovian picture where the light field is traced out. In addition, a rotating frame with the transition frequency ($\omega_0$) is chosen. As described in the companion paper, the equations of motion of a 1D waveguide can be written compactly as follows. First, we define the single atom constants $r_i = \sqrt{\betaf_i} e^{-\imu k_{1D} z_i}$ and $l_i = \sqrt{\betab_i} e^{+\imu k_{1D} z_i}$, where $k_{1D}$ is the wavenumber of the transition, and $\betaf_i$ ($\betab_i$) is the forward (backward) coupling of atom $i$ (taken to be the same for every atom in this work). For the chiral case, the positions of the atoms can be rotated away by a frame transformation \cite{TWA2024cascaded}, so the atomic positions do not affect the dynamics. For the permutationally symmetric case, the atoms are positioned in the mirror configuration $k_{1D}(z_j-z_i)=n\pi$ for an integer $n$ \cite{anaPhysRevLett.131.033605}. Then, we define the right ($\wgmode{j}$) and left ($\wgbmode{j}$) propagating modes impinging on the $\nth{j}$ atom 
\begin{equation}
    \wgmode{j} = \alpha_f - \imu \sum_{i<j} r_i \spindown{i},
\end{equation}
\begin{equation}
    \wgbmode{j} = \alpha_b - \imu \sum_{i>j} l_i \spindown{i},
\end{equation} where $\alpha_f$ and $\alpha_b$ are proportional to the forward and backward input fields (taken to be $0$ throughout this work), and $\spindown{n} = \ket{g_n} \bra{e_n}$ is the spin-lowering operator of the $\nth{n}$ atom.. The equation of motion then reads
    \begin{align}\label{eq:master_bidirectional}
        \ddt \density &= -\imu \sum_n \{ r_n^* [\spinup{n}, \wgmode{n} \density] + l_n^* [\spinup{n}, \wgbmode{n} \density] \} + \ha + \mathcal{L}_{ng}[\density],   
    \end{align}
where $\mathcal{L}_{ng}[\density] = \sum_n  \spindown{n} \density \spinup{n} - \frac{1}{2} \left\{ \spinup{n}\spindown{n}, \density \right\}$ and $\ha$ means the Hermitian adjoint of the previous terms. We also work in the units where $\Gamma_0 = 1$, so the time has units of the lifetime of the single-atom excitation ($1/\Gamma_0$). The commutator terms describe the interaction of the atoms with the collective modes of left and right atoms and are responsible for collective dynamics and correlations among atoms. $\mathcal{L}_{ng}$ describes the individual decay of atoms to freespace and the waveguide. We note that the compact form of the master equation in \Eq{\ref{eq:master_bidirectional}} is equivalent to the standard form with a coherent Hamiltonian and Lindblad dissipators in Ref.~\cite{TWA2024cascaded} and the companion paper \cite{tohfachiral}. The form employed here, however, explicitly reveals the collective nature of the setup, where an atom interacts with the collective modes of all of its left or right atoms.

The goal of this work is to find analytical predictions in the thermodynamic limit for time-dependent correlators of the output modes of the waveguide,
\begin{equation}\label{eq:output}
    \wgmode{\text{out}} = \wgmode{N+1}, \quad \wgbmode{\text{out}} = \wgbmode{N+1}.
\end{equation}
such as the right, left, and total output flux 
\begin{align}\label{eq:power_output}
    &P_r(t) = \expect{\wgmode{\text{out}}^\dagger(t) \wgmode{\text{out}}(t)}, \quad P_l(t) = \expect{\wgbmode{\text{out}}^\dagger(t) \wgbmode{\text{out}}(t)}, \nonumber \\
    &P(t) = P_r(t) + P_l(t),
\end{align} and the two-time second-order correlation
\begin{equation}\label{eq:G2t1t2}
    G^{(2)}(t_1,t_2) = \expect{\wgmode{\text{out}}^\dagger(t_1) \wgmode{\text{out}}^\dagger(t_2)  \wgmode{\text{out}}(t_2)\wgmode{\text{out}}(t_1)},
\end{equation} and the normalized correlation
\begin{equation}\label{eq:g2}
    g^{(2)}(t_1,t_2) = \frac{G^{(2)}(t_1,t_2)}{P_r(t_1)P_r(t_2)}.
\end{equation} For the chiral system, $\wgbmode{\text{out}} = 0$, and for the symmetric system, $\wgmode{\text{out}} = \wgbmode{\text{out}}$. The method and references for computing two-time correlators are mentioned in the companion paper \cite{tohfachiral}. An analytic solution of the master equation \Eq{\ref{eq:master_bidirectional}} for a chiral system based on a $\beta$ expansion was developed in the companion paper \cite{tohfachiral}, and is used here to explore the thermodynamic limit in Sec.~\ref{sec:analytical_methods}.


Because of the exponential computational complexity of the master equation \Eq{\ref{eq:master_bidirectional}}, we turn to efficient approximation methods. The main method is the mean field or the cumulant expansion. A mean field method of order $n$ keeps track of atomic operators involving up to $n$ atoms while approximating higher-order operators \cite{MFspinskramer2015generalized}. Such method is denoted MF-$n$ \cite{MFrobicheaux2021beyond}. As shown later, a second-order mean field method (MF2) is exact in the thermodynamic limit. This is in contrast with finite $N$ and non-zero $\beta$, where MF2 is an approximation, and higher order MF might be necessary \cite{tohfachiral}. We only use MF2 in the main text to describe the limit $N \rightarrow \infty$. For \emph{finite} $N$, corrections based on higher order MF are required as shown in App.~\ref{app:beyond_thermodynamic_limit}. The derivation of the MF2 equations and an efficient algorithm to propagate them for a finite $N$ are described in the appendix of the companion paper \cite{tohfachiral}.

In this work, we are interested in the $\beta \rightarrow 0$ and the continuum limit for a chiral system applies. The continuum limit of the MF2 equations and its solution is treated in Sec.~\ref{sec:continuum_approximation}. The corresponding limit for a symmetric system is treated in Sec.~\ref{sec:continuum_approximation_sym}.

\subsection{Analytical Methods} \label{sec:analytical_methods}

In the companion paper \cite{tohfachiral}, we outlined a method to extract an analytic $\beta$ expansion for the power radiated in the chiral waveguide, $P(t)$, as well as the second order correlators, $G^{(2)}(t,t)$ and $G^{(2)}(0,t)$, starting from full inversion and with homogeneous coupling. For example, $P(t)$ can be expanded as
\begin{align}\label{eq:analytical_power}
    P(t, N, \beta) &= \sum_{k=1} \beta^k f_k(N,t).
\end{align} In Ref.~\cite{tohfachiral}, the coefficient functions, $f_k(N,t)$, become increasingly intricate for a general $N$, and we could only obtain up to $k_{max} = 8$ due to computational difficulties. This is due to the multiple timescales contributing to the dynamics progressively. However, the functions, $f_k(N,t)$, admit a simpler structure in the thermodynamic limit $N \rightarrow \infty$ keeping the optical depth $OD = 4N\beta$ fixed. For ease of notation, we define a scaled optical depth
\begin{equation}\label{eq:B}
 \B = N\beta.
\end{equation}
For example, the radiated power, $P(t)$ in a chiral waveguide, becomes
\begin{align}\label{eq:analytical_power_th_limit}
    P(t, \B) &= \B e^{-t} \sum_{n=0} c_n \B^n h^n(t),
\end{align} where 
\begin{equation}\label{eq:h}
    h(t) = -(2e^{-t}+t-2),
   \end{equation}
and $\{c_n\}$ is the sequence of constants $\{1,\, 1,\, \frac{1}{2},\, \frac{5}{36},\, \frac{7}{288},\, \frac{7}{2400},\, \frac{11}{43,200},\, \frac{143}{8,467,200},\, \dots \}$. These coefficients up to the $\B^8$ order were explicitly calculated from the limit of the found $\beta^k f_k(N,t)$ and have the following recurrence relation
\begin{align}\label{eq:analytical_coeff_recurrence}
    \frac{c_{n+1}}{c_{n}} = \frac{2 (2n+1)}{(n+1)^2(n+2)}.
\end{align}
Interestingly, these coefficients can be found from the solution of the MF2 equations in the continuum limit as illustrated in Sec.\ref{sec:continuum_approximation}. This establishes the exactness of the continuum limit of MF2 and allows us to calculate many more coefficients or even obtain closed form solutions for the time dependence of several observables. Similar analysis can be done for $G^{(2)}(t,t)$ starting from full inversion as well as for $P(t)$ starting from the $\ket{\psi}_{N-1}$.

The analytical solution in \Eq{\ref{eq:analytical_power_th_limit}} reveals a special time, $t_{sp} \approx 1.59$, which is the zero of $h(t)$. This time marks the separation between superradiant and subradiant decay as shown later. We note that this special time also arises in a permutationally symmetric system as illustrated in Sec.~\ref{sec:continuum_approximation_sym}. We highlight several other key results in Sec.~\ref{sec:results_thermodynamic_limit}.
\subsection{Continuum approximation for a chiral system}\label{sec:continuum_approximation}
In this section, we develop a continuum approximation for the MF2 equations for a chiral system similar to what is used in Ref.~\cite{Retardation2025effects} for partially inverted systems using MF1. There, the totally inverted state is adynamical, which is not physical, suggesting the need for a MF2-based approximation to capture the build of coherence and enhanced emission. To this effect, we cast the equations of MF2 (see the companion paper \cite{tohfachiral}) as a set of coupled partial integro-differential equations. In the case of no drive starting from inversion, the only surviving terms are $\exc{l}\,, \spinup{l}\spindown{k}\,, \exc{l}\exc{k}$ which can be mapped to real functions of one or two variables $x$ and $y$. The continuum approximation is expected to work in the limit $\beta \rightarrow 0$ as the dynamics at a specific atom does not differ considerably from nearby atoms. In this view, the $\beta$ parameter serves as the discretization step $\Delta x = \Delta y = \beta$ of the `optical space'. This is because an atom at index $l$ acquires an optical depth $\propto l \beta \equiv x$. Specifically, the leftmost atom is at optical depth $x = 0$, and $x$ increases further downstream until the rightmost atom where $x=\B = N\beta$.


We map the expectation values of the indexed variables $\exc{l}\,, \spinup{l}\spindown{k}\,, \exc{l}\exc{k}$ into continuous functions as follows
\begin{align}\label{eq:mapping_discrete_continuous}
&\expect{\exc{l}} \leftrightarrow e(x = l\beta), \quad \expect{\spinup{l}\spindown{k}} \leftrightarrow C(x = l\beta, y = k\beta), \nonumber \\ &\expect{\exc{l}\exc{k}} \leftrightarrow E(x = l\beta, y =k \beta),
\end{align} with the time argument suppressed for brevity. With $\beta \rightarrow 0$, the equations of motion become

\begin{align}\label{eq:e_xt}
    (\partial_t + 1) \, e(x,t) &= -2 \int_{0}^{x} C(x, x', t) \, dx',
\end{align}

\begin{align}\label{eq:C_xyt}
    (\partial_t + 1) \, C(x,y,t) &= \left[ 2e(y,t) - 1 \right] \int_{0}^{y} C(x, y', t) \, dy' + (x \leftrightarrow y) \nonumber \\
    &\quad + \beta \left[ 2 E(x,y, t) - e(\min\{x,y\},t) \right],
\end{align}

\begin{align}\label{eq:E_xyt}
    (\partial_t + 2) \, E(x,y,t) &= -2e(x,t) \int_{0}^{y} C(x, y', t) \, dy' + (x \leftrightarrow y),
\end{align} where $\partial_t$ means derivative with respect to time, and $(x \leftrightarrow y)$ means the previous term with $x$ and $y$ exchanged. $\min\{x,y\}$ means the minimum of the two values. Note in this setting, both $C(x,y)$ and $E(x,y)$ are symmetric with respect to exchanging $x$ and $y$. This is because they start in a symmetric form and evolve under symmetric equations. Note that the term proportional to $\beta$ in \Eq{\ref{eq:C_xyt}} is essential. This is because it is responsible for sourcing the dominant order of the correlation, $C$, when starting from inversion. Without this term, we get the trivial independently decaying ensemble as shown below. 

Observables of interest can also be mapped to continuous functions. For example, the optical power emitted at depth $x$ can be represented as

\begin{align}\label{eq:P_xt}
    P(x, t) = \int_{0}^{x} e(x',t) \, dx' + D(x, t),
\end{align} where we introduced the doubly integrated correlator
\begin{align}\label{eq:D_xt}
    D(x, t) = \frac{\int_{0}^{x} \int_{0}^{x} C(x', y', t) dx' dy'}{\beta}.
\end{align} Note that the correlator $C$ is proportional to $\beta$ as shown below, so $D$ is well defined in the limit $\beta \rightarrow 0$. Similarly, the second order coherence $G^{(2)}(t,t)$ in \Eq{\ref{eq:G2t1t2}} at optical depth $x$ can be mapped to the function

\begin{align}\label{eq:Q_xt}
    Q(x, t) &=  2  \left[ P(x, t) \right]^2 \nonumber \\ &+ 2\int_{0}^{x} \int_{0}^{x} \left[ E(x', y', t) - e(x',t) e(y',t)\right] dx' dy',
\end{align} which is factorizable as $2 \left[ P(x, t) \right]^2$ without the second integral term. To zeroth order in $\beta$, the second term vanishes when starting from the excited state $\allexcited$, as shown below. Thus, the second order coherence $g^{(2)}(t,t) = 2$ for all optical depth in the thermodynamic limit. Therefore, the observed build up of the second order coherence in the experiment \cite{bach2024emergence} and in the companion paper \cite{tohfachiral} hinges on the finite-size nature of the system; i.e. $\beta$ needs to be non-zero even if assume that $N$ is going to infinity. The first order correction, $\mathcal{O}(\beta)$, for $g^{(2)}(t,t)$ for a \emph{symmetric} system is analyzed in App.~\ref{app:beyond_thermodynamic_limit}.


We are interested in the solution starting from the fully inverted state, for which we have the initial conditions
\begin{equation}\label{eq:init_cond_fully_inverted}
    e(x, 0) = 1, \quad C(x, y, 0) = 0, \quad E(x, y, 0) = 1.
\end{equation} The above mean field equations can be generalized to cases with varying coupling $\beta$ by introducing $\beta(x)$. In addition, they can be generalized to cases with a driving laser but would require more variables and more equations.


It is notable that a perturbative solution to the continuum MF2 equation yields exactly the same dynamics represented in the analytical formula in \Eq{\ref{eq:analytical_power_th_limit}}. A perturbative solution ansatz starting from full inversion exists in the form $F \approx F_0 + \beta F_1$, where $F$ stands for any of the variables $e(x)\,, C(x,y)\,, E(x,y)$. The zeroth order solution is the independently decaying ensemble with $e_0(x, t) = e^{-t} \,, C_0(x,y,t) = 0\,, E_0(x,y,t) = e^{-2t}$. Inserting these quantities in the equations of motion yields linear equations for $F_1$. For example, for $C_1$ we get  

\begin{align}\label{eq:C1_xyt}
    (\partial_t + 1) \, C_1(x,y,t) &= \left[ 2e^{-t} - 1 \right] \int_{0}^{y} C_1(x, y', t) \, dy' \nonumber \\ &+ (x \leftrightarrow y) + \left[ 2 e^{-2t} - e^{-t}  \right].
\end{align} We observe that the zeroth order excitation acts as source for the correlation. This integro-differential equation can be solved by a simple ansatz

\begin{align}\label{eq:C1}
   C_1(x,y,t) &= e^{-t} \sum_{i,j} c_{ij} x^i y^j h(t)^{i+j+1},
\end{align} where $h(t) = -(2e^{-t}+t-2)$ and $c_{ij}$ are constants that can be obtained iteratively

\begin{align}\label{eq:cij}
    c_{0,0} &= 1, \\
    c_{i,j} &= \frac{1}{i+j+1}(\frac{c_{i,j-1}}{j} + \frac{c_{i-1,j}}{i}), 
\end{align} with the exception for $i=0$, where $c_{0,j} = \frac{1}{j+1}(\frac{c_{0,j-1}}{j})$, and similarly for $j=0$. The formula \Eq{\ref{eq:C1}} has interesting implications regarding the correlation between the atoms at the peak radiation time, at the special time $t_{sp}$ and at late time $t > t_{sp}$. The influence on the excitation decay $e_1$ can also be found by integrating the correlation as in \Eq{\ref{eq:e_xt}}. These results are presented in Sec.~\ref{sec:corr_exc}.

In deriving the continuum MF2 equations, we replace discrete sums by integrals. For example, a quantity $\sum_i^{l-1} \expect{\spinup{k}\spindown{i}} \beta$ is turned into $\int_{0}^{y=l\beta} C(x = k\beta, y') \, dy'$. This incurs an error which is $\mathcal{O}(y \beta \sup[\partial_y \,C])$, where $\sup$ is the supremum function. However, $C$ is proportional to $\beta$. This implies that the error is $\mathcal{O}(\beta^2)$. There are several sources of error at this order, so in the main text, we only find solutions to order $\mathcal{O}(\beta)$, which become exact in the $\beta \rightarrow 0$ limit. It is notable that corrections to the thermodynamic limit (i.e., for finite $N$ or non-zero $\beta$) generally require going beyond MF2. This is shown for a symmetric system in App.~\ref{app:beyond_thermodynamic_limit}.


Having obtained the correlation $C$, the quantity $D$ in \Eq{\ref{eq:D_xt}} interestingly sums up to a simple formula based on a generalized hypergeometric function
\begin{align}\label{eq:D_xt_summed}
    D(x, t) = x e^{-t} \left[\,_1F_2\left(\frac{1}{2};1,2;4 x h(t)\right) - 1\right],
\end{align} from which the formula for the radiated power in the waveguide can be obtained 

\begin{align}\label{eq:power_hypergeometric}
    P(x, t) = x e^{-t} \,_1F_2\left(\frac{1}{2};1,2;4 x h(t)\right),
\end{align} which is equivalent to the series expansion in \Eq{\ref{eq:analytical_power_th_limit}}, with $x$ standing for $\B$. This formula is exact to zeroth order in $\beta$ and thus is exact in the limit $\beta \rightarrow 0$. Interestingly, this hypergeometric function satisfies the following relation
\begin{align}\label{eq:bessel_hypergeometric}
    \,_1F_2\left(\frac{1}{2};1,2;z\right) =  I_0(\sqrt{z})^2-I_1(\sqrt{z})^2,
\end{align} where $I_0$ and $I_1$ are modified Bessel functions of the first kind. The square root inside the argument of the Bessel functions indicates that the behavior of the system depends on the $\sqrt{xh(t)}$, which is known as a \emph{stretched} behavior. In other words, the effect of the optical depth is \emph{quadratically} suppressed. This is in contrast to the symmetric system as we show in Sec.~\ref{sec:continuum_approximation_sym}. The square root is also responsible for an oscillatory behavior at late times. Consequences of this analytical formula are investigated in Sec.~\ref{sec:results}, and its asymptotic behavior can be found in App.~\ref{app:asymptotic}.

In order to measure $g^{(2)}(0,t)$, we also need the power from the Dicke state $\ket{\psi}_{N-1} \propto \sum_n \spindown{n} \ket{ee\dots e}$. This corresponds to another solution of the system with the initial conditions 
\begin{equation}\label{eq:init_cond_lower_dicke}
    e(x, 0) = 1 - \frac{\beta}{\B}, \quad C(x, y, 0) = \frac{\beta}{\B}, \quad E(x, y, 0) = 1 - \frac{2\beta}{\B},
\end{equation}  where $\B$ is proportional to the optical depth of the array, $\B = N \beta$.

This specific initial state alters the expansion coefficients in \Eq{\ref{eq:cij}} by adding another term that has a slightly different recurrence relation and seed
\begin{align}\label{eq:dij}
    d_{0,0} &= -1, \\
    d_{i,j} &= \frac{1}{i+j}(\frac{d_{i,j-1}}{j} + \frac{d_{i-1,j}}{i}),
\end{align} with the exception for $i=0$, where $d_{0,j} = \frac{1}{j}(\frac{d_{0,j-1}}{j})$, and similarly for $j=0$. This change, however, prohibits us from finding a closed form for the $D$. Thus, we find the coefficients needed for calculations by numerically solving the recurrence relation. Nonetheless, this is an $\mathcal{O}(k_{max}^2)$ in complexity, where $k_{max}$ is the truncation order in a series expansion, and can be used easily to generate enough coefficients for convergence of calculations for $\B \sim 10^4$. We demonstrate the effect of the optical depth on $g^{(2)}(0,t)$ in Fig.~\ref{fig:g2_0t_therm_plot} in Sec.~\ref{sec:results_thermodynamic_limit}.






\subsection{Thermodynamic limit of a symmetric system}\label{sec:continuum_approximation_sym}
In this section, we solve the MF2 equations for the permutationally symmetric systems. The symmetric system occurs when the atoms are coupled symmetrically to both the left and right propagating photons and are positioned in the mirror configuration \cite{anaPhysRevLett.131.033605}. The equations of motion in MF2 are much simpler than in the chiral system resulting in much simpler dynamics. In this case, all atoms have identical dynamics, and the position dependence in the quantities $e(x)$, $C(x,y)$, and $E(x,y)$ is dropped.

It is of interest to compare the dynamics of the chiral and permutationally symmetric systems. First, the symmetric system can serve as a base point to which the more complicated chiral system can be compared. These systems differ only in the coherent Hamiltonian, and thus, have the same initial dynamics when starting from full inversion \cite{robicheaux2021theoretical,anaPhysRevLett.131.033605}, which can be used as a sanity check to validate the derived equations of motion. Second, this comparison illuminates the role of chirality or a broken symmetry on the complexity and the physics of the collective light-atom interaction. Further, it is possible to mathematically prove the exactness of MF2 for a symmetric system as we show in App.~\ref{sec:proof}.


For the symmetric system, the finite $N$ equations can be obtained via the package in Ref.~\cite{juliaplankensteiner2022quantumcumulants}, then the infinite $N$ limit is obtained. With $\beta \rightarrow 0$ and $\beta N = \B$, the equations of motion become

\begin{align}\label{eq:e_xt_sym}
    (\partial_t + 1) \, e(t) &= -\B C(t),
\end{align}

\begin{align}\label{eq:C_xyt_sym}
    (\partial_t + 1) \, C(t) &= \B \left[ 2e(t) - 1 \right] C(t) + \beta \left[ 2 E(t) - e(t) \right], 
\end{align}

\begin{align}\label{eq:E_xyt_sym}
    (\partial_t + 2) \, E(t) &= -2 \B e(t) C(t).
\end{align} The total power emitted in both directions of the waveguide in this case is
\begin{align}\label{eq:P_xt_sym}
    P(t) = \B e(t) + \frac{\B^2 C(t)}  {\beta},
\end{align} which in turn can be shown to obey a simple homogeneous ODE when the system starts from a Dicke state with a finite number of ground state atoms such that $e(0) = 1 - \mathcal{O}(\beta)$ and $e(0) \rightarrow 1$ when $\beta \rightarrow 0$. This ODE is

\begin{align}\label{eq:P_xt_sym_ODE}
    (\partial_t + (\B + 1 - 2\B e^{-t})) P(t) = 0.
\end{align} We could not find an analog of this homogeneous ODE for the chiral system which has a more complicated power formula. Equation (\ref{eq:P_xt_sym_ODE}) has a simple solution

\begin{align}\label{eq:P_xt_sym_soln}
    P(t) = P(0) e^{-t} \exp \{ \B h(t)\},
\end{align} where $h(t)$ is from \Eq{\ref{eq:h}}. Equation (\ref{eq:P_xt_sym_soln}) leads to trivial second order correlations $g^{(2)}(0,t) = 2$ starting from the $\ket{\psi}_{N}$ state. This is because $P(t)$ starting from both state $\ket{\psi}_{N}$ and $\ket{\psi}_{N-1}$ have the same form except for an initial factor of $2$. This is in contrast with the $g^{(2)}(0,t)$ for the chiral system, which is non-trivial. Additionally, $g^{(2)}(t,t) = 2$ because of the same factorization noted for the chiral system in Sec.~\ref{sec:continuum_approximation}. The special time $t_{sp} \approx 1.59$ also emerges in this system and separates super- and sub-radiance. In Sec.~\ref{sec:results}, we plot the power, $P(t)$, for the symmetric and chiral systems to compare the radiation dynamics.

The total energy emitted in both directions of the waveguide, $E_{wg}$ from the totally inverted state (with $P(0) = \B/2$) can be obtained by integrating $P(t)$ in \Eq{\ref{eq:P_xt_sym_soln}} resulting in
\begin{align}\label{eq:energy_wg_sym}
    E_{wg} = \frac{e^{2\B}}{2(2\B)^\B}\gamma(\B+1, 2\B) \sim \sqrt{\frac{\pi \B}{2}} (\frac{e}{2})^\B,
\end{align} where $\gamma$ is the lower incomplete gamma function and the last asymptotic behavior holds for large $\B$ and is based on the Stirling's approximation. The energy here is in units of the atomic transition or the emitted photon energy. This shows an exponential enhancement in the energy radiated into the waveguide in the thermodynamic limit. As shown in Sec.~\ref{sec:results_thermodynamic_limit}, this implies that the minimum $N$ to achieve the thermodynamic limit also grows exponentially with the optical depth. In a different thermodynamic limit, where $N \rightarrow \infty$, but with \emph{fixed} $\beta$, it is shown in Ref.~\cite{NlimitPhysRevA.106.013716} that $E_{wg} \rightarrow N$, so that \emph{almost all} the energy is emitted into the waveguide.




\section{Results} \label{sec:results}
In this section, we present results pertaining to photon statistics in the chiral and symmetric waveguide system in the thermodynamic limit. First, we show the approach to the thermodynamic limit using both the analytical and MF2 approximations. We demonstrate that they converge to an exact answer presented in Sec.~\ref{sec:continuum_approximation} for the chiral system. Second, we show the behavior of the photon rate and the photon-photon correlations in the thermodynamic limit. Finally, we study the correlation development and the enhanced excitation decay for a chiral system.


\subsection{Approaching thermodynamic limit at fixed $N\beta$}\label{sec:approaching_therm_limit}

\begin{figure}[htbp]
    \centering
    \includegraphics[width=0.45\textwidth]{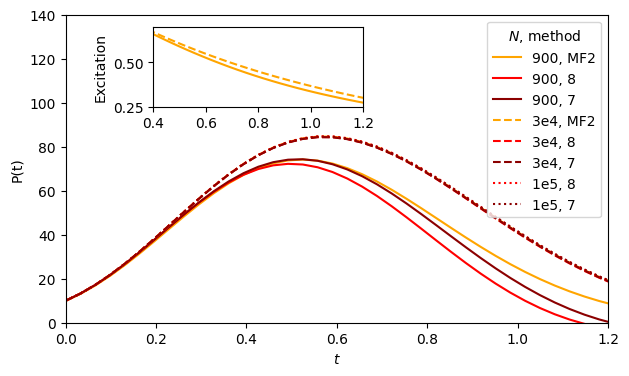}
    \caption{\justifying Comparison of the dynamics of the power radiated into a chiral waveguide starting from the inverted state for the MF2 approximation (orange), and the analytical solutions of truncation order ($k_{max}$) of 8 (red) and 7 (dark red) for $N \in \{ 900, 3 \times 10^4, 1 \times 10^5\}$ in solid, dashed, and dotted lines respectively. The inset shows the excitation for MF2.}
    \label{fig:thermodynamic_limit}
\end{figure}

In this section, we explore the approach to the thermodynamic limit $N \rightarrow \infty$, while keeping the $N \beta \equiv \B$ fixed. We observe that the dynamics of different observables reach a limiting behavior as predicted by the discussions in Secs.~ \ref{sec:analytical_methods} and \ref{sec:continuum_approximation}. For example, in Fig.~\ref{fig:thermodynamic_limit}, we fix $\B = 10$ for a chiral system. The limit is reached when $N \gtrsim 3 \times 10^4$ when we look at the power emitted in the waveguide from the $\allexcited$ state in the time window $t < 1.2$. For larger $\B$ or longer time window, the $N$ required to reach the limit is higher. Additionally, when the limit is reached, we find that MF2 agrees perfectly with the converging analytical solution which admits the analytical form in \Eq{\ref{eq:power_hypergeometric}}. This establishes the exactness of the MF2 method and hints at the system reaching a relatively simple behavior in the thermodynamic limit. Another evidence for the exactness of MF2 comes from the agreement of the expansion coefficients of the predicted power formula in \Eq{\ref{eq:power_hypergeometric}} with the first eight explicitly calculated coefficients from the solution of the full dynamics in \Eq{\ref{eq:analytical_power_th_limit}}. In App.~\ref{sec:proof}, we provide a mathematical proof for the exactness of MF2 for a \emph{symmetric} system.


While the thermodynamic limit is an idealization, large $N \sim 900$ cases have been experimentally demonstrated \cite{bach2024emergence,PRXsuperradiant2024bursts}. However, as seen in Fig.~\ref{fig:thermodynamic_limit}, an $N = 900$ setup behaves similar to its thermodynamic limit, which is much easier to compute using the analytical formulas provided. Thus, the thermodynamic limit can give quick insight into the behavior of future experiments.

\begin{figure}[htbp]
    \centering
    \includegraphics[width=0.45\textwidth]{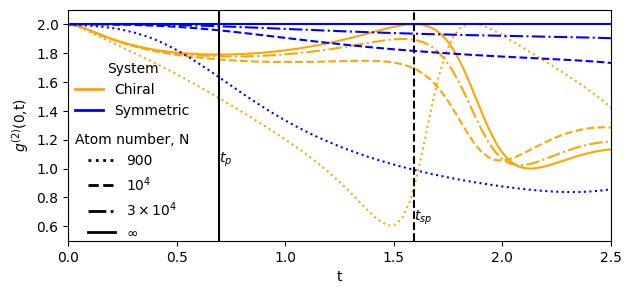}
    \caption{\justifying Comparison of the dynamics of the second-order correlation $g^{(2)}(0,t)$ starting from the inverted state for the MF2 approximation for a chiral system(orange), and a permutationally symmetric system (blue). $N \beta \equiv \B = 10$ and  for $N \in \{ 900, 1 \times 10^4, 3 \times 10^4, \infty\}$. The infinite $N$ curves (in solid) come from the analytical formulas, while the other finite $N$ curves were numerically integrated.}
    \label{fig:thermodynamic_limit_g2_0t}
\end{figure}

We can also look at the second-order correlation $g^{(2)}(0,t)$ which relates to initial shot-to-shot noise and intensity fluctuation in the emitted photon rate. In Fig.~\ref{fig:thermodynamic_limit_g2_0t}, we fix $\B = 10$ and increase $N$. We observe that smaller $N$ cases feature more variation in $g^{(2)}(0,t)$ and thus more intensity fluctuation. The variation is minimized in the thermodynamic limit $N \rightarrow \infty$. Interestingly, in this limit, the permutationally symmetric system features no variation in $g^{(2)}(0,t)$, while the chiral system has some noticeable variation especially at $t=0$ and the special time $t_{sp} \approx 1.59 /\Gamma_0$.

\begin{figure}[htbp]
    \centering
    \includegraphics[width=0.45\textwidth]{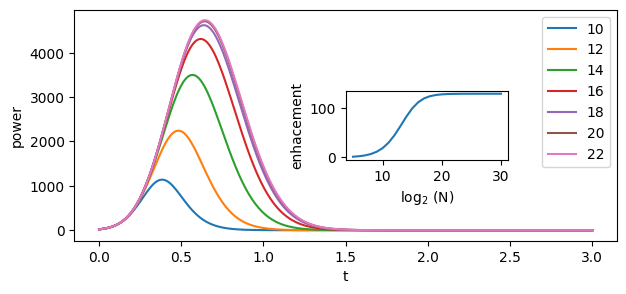}
    \caption{\justifying MF2 prediction of the power emitted in both directions of a symmetric waveguide starting from the fully inverted state. $\B = N \beta$ is fixed at $20$, while $N$ is increased until the thermodynamic limit is reached at $N \sim 2^{20}$. Legend shows $\log_2(N)$. Inset shows the enhancement factor of the total energy radiated in the waveguide due to collective effects.}
    \label{fig:thermodynamic_limit_symmetric}
\end{figure}

Because we are studying the limit at fixed $\B = N \beta$, the thermodynamic limit is only approached at very small coupling $\beta$. To illustrate, for a symmetric system with $\B = 20$, the thermodynamic limit is reached for $N \sim 2^{20}$. This implies that $\beta \lesssim 2 \times 10^{-5}$. However, due to collective enhancement, a fraction of the initial energy stored in the system much bigger than $\beta$ is emitted into the waveguide. This is illustrated in Fig.~\ref{fig:thermodynamic_limit_symmetric}.

Figure \ref{fig:thermodynamic_limit_symmetric} shows the power emitted into both direction of the waveguide according to MF2 for $\B = 20$ as the thermodynamic limit is approached. We observe that the power increases as $N$ increases until the thermodynamic limit is reached at $N \sim 2^{20}$. The power stays constant for larger $N$. The total energy going into the waveguide is given by the integral of the power. As the thermodynamic limit is reached, this energy increases and corresponds to $\approx 2600$ photons for $N = 2^{20}$. This is a tiny fraction of the initially stored energy corresponding to $2^{20}$ photons and is approximately $0.26 \%$. However, it is greatly enhanced from independent emission which would result in $\B = 20$ photons emitted in the waveguide. That is a factor of $\approx 130$ improvement as shown in the inset. If we had chosen a smaller atom number, $N = 2^{14}$, the improvement factor would be smaller ($\approx 83$) according to MF2. However, this factor is only an approximation because $N = 2^{14}$ is below the thermodynamic threshold $N \sim 2^{20}$, and MF2 is not exact before getting to the limit. Nevertheless, the fraction of photons emitted into the waveguide for $N = 2^{14}$ is greater than in the thermodynamic limit and is $\approx 10\%$. Again, this is a greatly enhanced fraction compared to that from independent emission (which is $\beta \approx 0.12\%$).


For practical experiments, $N$ is finite, and approximating the system by its thermodynamic limit incurs an error. We find the error to be small as long as a small fraction, ($\lesssim 1\%$), of the total energy is captured by the waveguide. For example, when this energy fraction condition is satisfied, the relative error between the peak photon rate for finite $N$ and infinite $N$ is ($\lesssim 8\%$). This was checked for a \emph{symmetric} system for a range of $\B$ from $0$ to $45$. We believe a similar case holds for a \emph{chiral} system, but this was not checked.




\subsection{Scaling with $\B$ in the thermodynamic limit}\label{sec:results_thermodynamic_limit}

It was shown in Sec.~\ref{sec:continuum_approximation} and Sec.~\ref{sec:analytical_methods} that the system acquires an exact analytical solution in a thermodynamic limit as $\beta \rightarrow 0$ keeping $\B = N \beta$ finite. In this section, we explore the behavior of the analytical solution when $\B$ is increased. First we define a normalized decay rate $\Gamma(x,t)$ that captures the enhancement in the decay rate relative to an independent ensemble. This normalized decay rate is obtained by dividing the power emitted into the waveguide by the energy stored in the array $\int_{0}^{\B} e(x,t) \, dx/\beta = N e^{-t} + \mathcal{O}(1)$ and scaling by the coupling $\beta$.  This is proportional to the definition in Ref.~\cite{oscillationPhysRevLett.128.203601}. Equivalently, this is the power emitted into the waveguide divided by the power from an independent ensemble, $P^{ind}(x,t)$, which is $\B e^{-t}$.

\begin{align}\label{eq:Gamma_definition}
    \Gamma(\B,t)  &= \frac{P(\B,t)}{\beta N e^{-t}} = \frac{P(\B,t)}{P^{ind}(\B,t)}.
\end{align}

\begin{figure}[htbp]
    \centering
    \includegraphics[width=0.45\textwidth]{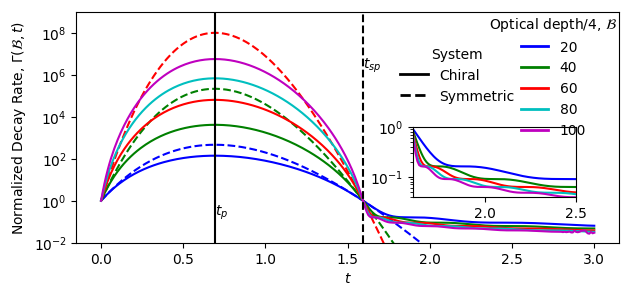}
    \caption{\justifying Normalized decay rate into the waveguide, $\Gamma(\B,t)$, from an ensemble of a given optical depth $\B$ for a chiral system (solid lines) and a symmetric system (dashed lines). The $\B = 80, \, 100$ curves for a symmetric system grow out of figure bound and are not shown.  Inset shows an oscillatory behavior in the radiation at late time for a chiral system, a signature of going through subradiant manifolds of the Hilbert space \cite{oscillationPhysRevLett.128.203601}.}
    \label{fig:exponential_power_growth}
\end{figure}

We find evidence of exponential enhancement in the collective behavior of the atoms as depicted by the peak decay rate into the waveguide in Fig.~\ref{fig:exponential_power_growth}. Additionally, at the peak decay time, a symmetric system is more superradiant than a chiral system. This is because the chiral system has coherent dynamics that tends to drive the state away from the maximally radiating or cooperative Dicke states.

Interestingly, the normalized decay rate, $\Gamma(\B,t)$, peaks at time $t_p = \ln(2)$ for both systems and for all optical depth $\propto \B$. This time comes from setting $h'(t) = 0$. Subsequently, for a symmetric system the peak normalized decay rate is 
\begin{align}\label{eq:G_max_sym}
    \Gamma_{max}^{sym}(\B)  &= (\frac{e}{2})^\B,
\end{align} which shows an exponential enhancement with the optical depth.

The time for peak \emph{decay rate}, $t_p$, is different from the peak time for the \emph{radiated photon rate} in \Eq{\ref{eq:power_hypergeometric}}. The latter is $0$ for $\B = 0$ (independently decaying ensemble) and shifts to $t_p = \ln(2)$ at large $\B$. Thus, in the large $\B$ limit, the peak photon rate is simply $P(\B, t_p)$. For a symmetric system, the peak radiation happens at $t = \ln(\frac{2\B}{\B+1})$ for $\B > 1$ and at $t = 0$ for $\B < 1$. Thus, the maximum power for the symmetric system, $P_{max}^{sym}$, for $\B > 1$ is 

\begin{align}\label{eq:p_max_sym}
    P_{max}^{sym}(\B)  &= B (\frac{\B+1}{2\B})^{\B+1} e^{\B-1} \sim \frac{\B }{2} (\frac{e}{2})^\B,
\end{align} where the last similarity holds for large $\B$ and indicates an exponential enhancement of the peak radiation power similar to the peak \emph{energy} in \Eq{\ref{eq:energy_wg_sym}}. For a chiral system, the enhancement is weaker, and is a \emph{stretched} exponential ($\B$ occurs under a square root in the exponent, $e^{4\sqrt{\B}}$) as shown in App.~\ref{app:asymptotic}.

The exponential growth of $P_{max}^{sym}(\B)$ can be used to set a lower bound on $N$ to achieve the thermodynamic limit. The maximum power into the waveguide is upper bounded by the system set in the state in the middle of the symmetric Dicke ladder. In this state, the radiated power is $P \sim \B N /4$. By setting $P_{max}^{sym}(\B) < \B N /4$, the minimum $N$ to achieve physicality can be found. For example, for $\B = 20$, $N$ must be greater than $\sim 950$. Moreover, this minimum $N$ increases \emph{exponentially} with increasing $\B$.



We also note the emergence of a special time $t_{sp} = 1.59$. Before $t_{sp}$, the emission into the waveguide is superradiant (normalized decay rate $>1$) and exponentially enhanced relative to an independently decaying ensemble when $\B$ is increased. In contrast, at late time, $ t > t_{sp}$, the system goes through subradiant states with normalized decay rate $<1$. For a symmetric system, such subradiance is \emph{exponentially} enhanced as $\B$ is increased. For example, in the subradiant region, the system has much smaller normalized decay rate into the waveguide for $\B = 40$ than for $\B = 20$. For a chiral system, the subradiant behavior is only \emph{algebraically} enhanced with increasing $\B$ (see App.~\ref{app:asymptotic}). Furthermore, the subradiant regime in a \emph{chiral} system is associated by an oscillatory behavior of the radiation. This oscillatory behavior has a progressively increasing period and was measured in the experiment \cite{oscillationPhysRevLett.128.203601} in the single-excitation manifold. We characterize this oscillation mathematically in App.~\ref{app:asymptotic}. Note that for both systems, as the optical depth increases, both super- and subradiance to the waveguide are more pronounced. In other words, as $\B$ increases, the normalized emission rate, $\Gamma(\B,t)$, increases before $t_{sp}$. After $t_{sp}$, as $\B$ increases, the normalized emission rate, $\Gamma(\B,t)$, \emph{decreases}.



\begin{figure}[ht]
    \centering
    \includegraphics[width=0.4\textwidth]{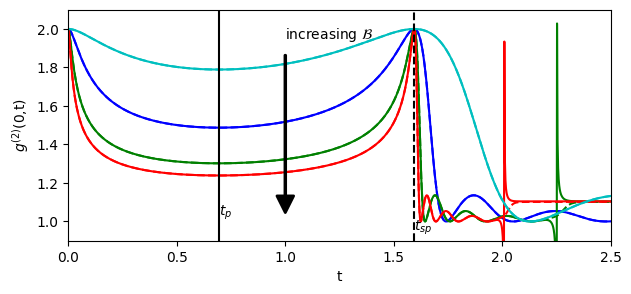}
    \caption{$g^{(2)}(0,t)$ for different values of $\B = 10, 40, 120, 200$. We use a cut off of 35 (solid) and 36 (dashed) in the series expansion of the numerator of $g^{(2)}(0,t)$ to show convergence. The $\B = 120$ (green) and  $\B = 200$ (red) curves are not converged for $t>2.25$ and $2.0$ respectively.}
    \label{fig:g2_0t_therm_plot}
\end{figure}

For a chiral system, the special time $t_{sp} = 1.59$ also separates two behaviors of the second order correlation $g^{(2)}(0,t)$ as depicted in Fig.~\ref{fig:g2_0t_therm_plot}. Before $t_{sp}$, the normalized correlation is mostly flat showing a decrease in the fluctuations of the intensity of the radiated photons. After $t_{sp}$, $g^{(2)}(0,t)$ drops to $1$ and oscillates near $1$. This also shows a decrease in the fluctuations of the intensity of the radiated photons. Most of the fluctuations happen at $t = 0$ and $ t = t_{sp}$, and they tend to grow with increasing $\B$. We also note that more terms in the series expansion of the $g^{(2)}(0,t)$ are required for convergence as $\B$ is increased especially at late times. This is depicted by some lack of convergence for $\B = 120, 200$ for $t > 2$ and the absence thereof for smaller $\B = 10, 40$. We note the contrasting behavior for a symmetric system, where $g^{(2)}(0,t) = 2$ for all $t$ and $\B$ in the thermodynamic limit as shown earlier in Fig.~\ref{fig:thermodynamic_limit_g2_0t}.

\subsection{Correlation development and excitation decay in the chiral system}\label{sec:corr_exc} 
In Sec.~\ref{sec:continuum_approximation}, we mathematically characterized the solution to the correlation development across the array, $C(x,y,t)$, for a chiral system. In this section, we highlight some key characteristics of this correlation. Additionally, we study the influence on the excitation decay as given by the first-order $\mathcal{O}(\beta)$ correction, $e_1$.

\begin{figure}[ht]
    \centering
    \includegraphics[width=0.4\textwidth]{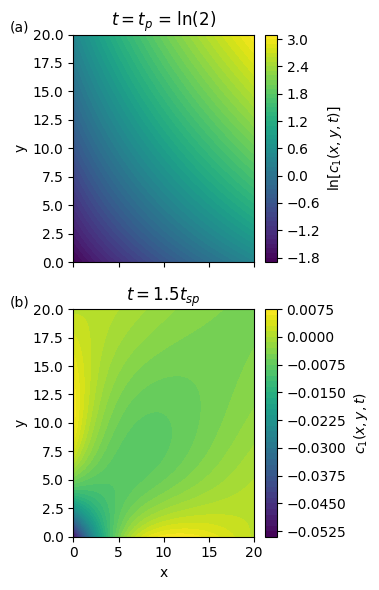}
    \caption{\justifying Development of the correlation $C_1(x,y,t)$ at various times of interest $t_p$ (a) and $1.5t_{sp}$ (b). Note the log scale for $t_p$. At $t_{sp}$ (not shown), the correlation is identically $0$ for all pairs $(x,y)$.}
    \label{fig:correlation}
\end{figure}

Because the correlation $C(x,y,t)$ determines the photon rate as in \Eq{\ref{eq:P_xt}} and subsequently the normalized decay rate as in \Eq{\ref{eq:Gamma_definition}}, it is of interest to investigate $C$ at specific times: the time of maximum decay rate $t_p$, the special time separating super- and subradiance $t_{sp}$, and a time in the late subradiance decay, chosen to be $1.5t_{sp}$. The correlation at $t_p$ and $1.5t_{sp}$ is a shown in Fig.~\ref{fig:correlation}. The correlation is identically $0$ at $t_{sp}$ (not shown in Fig.~\ref{fig:correlation}). At the maximum superradiance time $t_p$, most of the correlation is for nearby atoms at the downstream edge of the array (i.e. for large $x$ and $y$). This can be explained as follows. Two atoms at the upstream edge of the array radiate almost independently because we work in the limit of the coupling $\beta \rightarrow 0$. Further, they are not affected by the downstream atoms, resulting in small correlation developing between them. This is in contrast with two atoms downstream the chain which are irradiated by the same many upstream atoms resulting in large correlation developing between them.

Correlation between the left-most atom at $x=0$ and other atoms at the peak decay rate is captured by $C(0,y,t_p)$. Interestingly, the left-most atom is most correlated with the furthest atom at the right edge. Interestingly, this behavior was also seen in the \emph{steady} state of a coherently driven array in Ref.~\cite{MFchiral2023higherorder}.
    
This can be intuitively understood as follows. The left-most atom irradiates all the intermediate atoms which then act as a source for the right-most atom. This results in the correlation growing with the distance between atoms because there are more intermediate atoms mediating a long-range correlation. Mathematically, this correlation can be shown to be
\begin{align}\label{eq:corr_0y_tp}
    C_1(0,y,t_p) &= \frac{1-\ln(2)}{2} \frac{I_1(2\sqrt{s})}{\sqrt{s}} \sim \frac{1-\ln(2)}{2}\frac{e^{2\sqrt{s}}}{2\sqrt{\pi}s^{3/4}},
\end{align} where $I_1$ is a modified Bessel function of the first kind, $s = y[1-\ln(2)]$, and the last asymptotic behavior holds for large $s$. This shows a \emph{stretched} exponential behavior in the growth of the correlation with the `optical' distance between the atoms.


At $t_{sp}$, the correlation interestingly resets to $0$ between all atoms. This, however, does not imply that the atoms are in a product state to order $\beta$. This is because the cumulant ($E(x,y)-e(x)e(y)$) which measures the excitation correlation is not $0$ for all $x$ and $y$ to order $\beta$ at $t_{sp}$. (In contrast, for a symmetric system, $E(t_{sp})-e(t_{sp})e(t_{sp}) = 0$ to order $\beta$, meaning the atoms are in a product state at this order.)
At $t_{sp}$, the excitation $e(x,t) = e^{-t_{sp}}$. Afterwards, the system goes into a subradiant behavior where the correlation becomes negative for most pairs of atoms. This is shown in the bottom panel of Fig.~\ref{fig:correlation} at $1.5t_{sp}$. Interestingly, most of the anticorrelation happens between nearby atoms \emph{upstream} the chain. Further, the magnitude of the correlation tends to die out for far-away pairs, or nearby pairs \emph{downstream} the chain.

The amount of first order correlation $C_1(x,y,t)$ sets an upper bound on $\beta$ (and thereby a lower bound on $N$ if $\B$ is fixed) for which the continuum MF2 is valid. Because the correlation $ C = \beta C_1$ is physically bounded, $\frac{-1}{2} \leq C \leq \frac{1}{2}$, $\beta$ needs to be smaller than $\frac{1}{2C_1}$. For example, in in Fig.~\ref{fig:exponential_power_growth}, at $(x,y, t) = (20,20, t_p)$, $C_1 \approx 20$ and thus $\beta$ must be smaller than $1/40$ (thus, $N \gtrapprox 800$) for physicality. In practice, $\beta$ must be much smaller for the continuum approximation to hold.

Having analyzed the correlation development, the influence on the excitation decay $e_1$ can be found by integrating the correlation as in \Eq{\ref{eq:e_xt}}. This involves integrating powers of the function $h(t)$. We could not find a simple compact formula for such integrals, but $e_1$ can be represented as a series

\begin{align}\label{eq:e1_expanded}
    e_1(x,t) &= -e^{-t} \sum_{i=1} u_i x^i S_i(t),
\end{align} where, to aid the notation, we define the integrals
\begin{align}\label{eq:S_integrals}
    S_n(t) = \int_{0}^{t} h(t')^n dt' = \int_{0}^{t} [-(2e^{-t'}+t'-2)]^n dt'.
\end{align} $u_i$ can be obtained iteratively, with the first few nonzero coefficients are $\{2,\, \frac{3}{2},\, \frac{5}{9},\, \frac{175}{144},\, \dots \}$.
The first two $S_n$ integrals are
\begin{align}
    -2S_1(t) &= (t^2 - 4t + 4) - 4e^{-t}, \\
    -3S_2(t) &= 12(t - 1)e^{-t} + (-t^3 + 6t^2 - 12t + 6) + 6 e^{-2t}. 
\end{align} Similar to the verification of the exactness of the MF2 output power, $P(x,t)$ illustrated in Sec.~\ref{sec:analytical_methods}, we verified the first few coefficients of $e_1(x,t)$ coming from the continuum MF2 against the full solution of the Lindblad master equation \Eq{\ref{eq:master_bidirectional}} and taking the $N \rightarrow $ limit.

Note that the expansion is in the optical depth $\propto x$, suggesting the need for more terms as $x$ is increased. This intuitively agrees with the cascaded picture in which an atom is influenced by all atoms upstream. Thus, the dynamics of downstream atoms with large optical depth are more complicated than upstream atoms with small optical depth. In particular, the leftmost atom with $0$ optical depth is not affected and decays exponentially $e_1(0,t)=0$.

\begin{figure}[ht]
    \centering
    \includegraphics[width=0.4\textwidth]{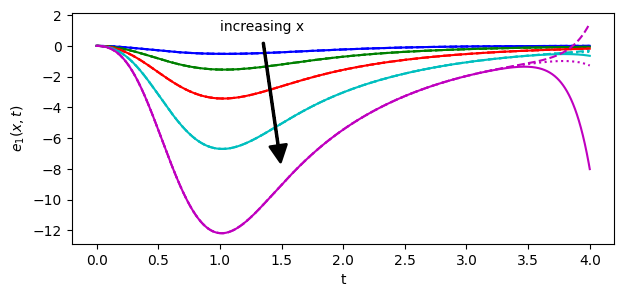}
    \caption{\justifying First order effect $e_1(x,t)$ on the excitation at different optical depths, where $x = 2, 4, 6, 8, 10$, as indicated by the arrow. We use a cut off in the sum in \Eq{\ref{eq:e1_expanded}} of 14 (solid lines), 15 (dashed lines), and 16 (dotted lines) to show convergence.}
    \label{fig:first_order_excitation}
\end{figure}

To illustrate the effects at various optical depth, we plot $e_1(x,t)$ for $x = 2, 4, 6, 8, 10$ in Fig.~\ref{fig:first_order_excitation}. We observe that as the optical depth is increased, the effect on the excitation increases exponentially. At early time, an excitation downstream tends to decay faster than an excitation upstream because of the increased incident radiation from the left (see schematic in Fig.~\ref{fig:intro_figure}). At late times, this effect diminishes as the atoms upstream have already decayed. We also note that for a higher optical depth, $e_1$ tends to have a minimum at a slightly earlier time. Additionally, as the optical depth, $x$, is increased, more terms in \Eq{\ref{eq:e1_expanded}} are required for convergence especially at late times. 

While in the thermodynamic limit, the first order correction $e_1$ is irrelevant, it comes useful in approximating the large $N$ cases. This is because at large enough but still finite $N$, the system's behavior becomes nearly identical to $N = \infty$ as we have showed in Sec.~\ref{sec:approaching_therm_limit}.

Similar to the argument for physicality of $C$, an argument for the excitation, $e$, can be made. Since $e(x,t)$ is non-negative, the $x = 10$ curve (for example) implies that $\beta \lessapprox 0.03$ for physicality of $e(x=10, t=1)$. In practice, $\beta$ must be much smaller for the thermodynamic limit to hold. For example, if $x = 10$, the system reaches the thermodynamic limit at $N \gtrsim 3 \times 10^4$, so that $\beta \sim 0.0003$.

\section{Conclusion}
\label{sec:conclusion}
In this work, we explored the thermodynamic limit $N \rightarrow \infty$ with the optical depth $OD = 4 N\beta$ fixed for an inverted array of atoms coupled to a waveguide. We explored a permutationally symmetric and a symmetry-lacking chiral systems and highlighted the differences both in complexity and collective radiation statistics. For both systems, the radiated power in the waveguide grows exponentially with the optical depth, and the same-time second-order coherence vanishes. However, the shot-to-shot fluctuation in photon rates is only trivial in the symmetric systems and shows interesting features for a chiral system.

It is of interest to explore an intermediate system where the coupling to right and left propagating modes are different and nonzero both for a mirror configuration and a general lattice constant \cite{anaPhysRevLett.131.033605}. For a system with a general lattice constant, dynamical mirror symmetry breaking occurs, and it is of interest to explore this behavior in the thermodynamic limit.

Data plotted in the figures is available at \cite{data}.

\begin{acknowledgments}
     We thank Wenqi Tong and AbdAlghaffar Amer for useful discussions and suggestions. This work was supported by the National Science Foundation under Award No. 2410890-PHY. This research was supported in part through computational resources provided by Information Technology at Purdue University, West Lafayette, Indiana.
\end{acknowledgments}

\appendix
\section{asymptotic behavior of $P$ for a chiral waveguide}\label{app:asymptotic}
In this appendix, we analyze the asymptotic behavior of $P(x,t)$ in \Eq{\ref{eq:power_hypergeometric}} to extract maximum power information as well as late time behavior. First, to simplify notation we denote $\,_1F_2\left(\frac{1}{2};1,2;4 x h(t)\right) \equiv R(4s)$, where $s = x h(t)$. The function $h(t) = -(2e^{-t}+t-2)$ in the argument is positive for $t < t_{sp}$ and negative for larger $t$. To find the peak radiation behavior, we find the asymptotic value for $R(4s)$ in the $s \rightarrow +\infty$ limit.
\begin{align}\label{eq:R_infty}
    R(4s) \sim \frac{e^{4\sqrt{s}}}{8\pi s}.
\end{align} This formula implies a stretched exponential enhancement of the radiation power with the optical depth.

For large $t$, we find the asymptotic value for $R(4s)$ in the $s \rightarrow -\infty$ limit.
\begin{align}\label{eq:R_infty}
    R(4s) \sim \frac{1}{\pi \sqrt{-s}} [1 - \frac{1}{4 \sqrt{-s}} \cos(4 \sqrt{-s})].
\end{align} This formula implies that for large enough $t$ such that $h(t) \sim 2-t$ and for large optical depth $\propto x$, the power radiated in the waveguide is 

\begin{align}\label{eq:R_neg_infty}
    P(x, t) &\sim x e^{-t} \frac{1}{\pi \sqrt{x(t-2)}} \nonumber \\
    &\times \left[1 - \frac{1}{4 \sqrt{x(t-2)}} \cos(4 \sqrt{x(t-2)})\right].
\end{align} This shows an algebraically modified decay dressed by a damped oscillatory behavior with progressively increasing period as depicted in Fig.~\ref{fig:exponential_power_growth}.
\section{Proof of exactness of MF2 for a symmetric system}\label{sec:proof}
In this appendix, we derive the equations for the moments of the symmetric system in the case of decay from an inverted state, and prove the exactness of the MF2 in the thermodynamic limit. First, we define notation for the nonzero moments. These are moments with equal number of spin raising ($\spinup{i}$) and lowering ($\spindown{i}$) operators, and any number of excitation operators $\exc{i}$. For example, $\spinup{i}\spindown{j}\exc{k}\exc{l}$ is nonzero during the evolution. Other operators with unequal number of spin raising and lowering operators is zero initially and remain zero throughout the evolution.

We denote the number of raising operators in a moment by $c$ and the number of excitation operators by $p$. For example, for $\spinup{i}\spindown{j}\exc{k}\exc{l}$, we have $c = 1$ and $p = 2$. We denote the expectation value of such operator by a 2-indexed variable $A_{p,c}$. With $N$ the number of atoms in the system, and the shorthand notation $N_i = N - i$ and $o = p + 2c$, the equation of motion for a generic $A_{p,c}$ reads 

\begin{align} \label{eq:generic_moment_sym}
    (\partial_t + p + c) A_{p,c} &= -c N_{2c} \beta A_{p,c} \nonumber \\
    &+ N_{o} \beta (2c A_{p+1,c} - p A_{p-1,c+1}) \nonumber \\
    &+ c^2 \beta (2 A_{p+2,c-1} - A_{p+1,c-1}),
\end{align} where for the inverted state, $\allexcited$, we have the following initial conditions    
\begin{equation} \label{eq:generic_moment_sym_initial}
A_{p,c}(0) = \delta_{c,0} =
\begin{cases}
1 & \text{if } c = 0 \\
0 & \text{if } c > 0.
\end{cases}
\end{equation}

In the thermodynamic limit, we have $\beta \rightarrow 0$ and $N \rightarrow \infty$. We thus define $\B = N\beta$, and do a perturbation analysis using the infinitesimal parameter $\beta$. First we rewrite \Eq{\ref{eq:generic_moment_sym}} as 
\begin{align} \label{eq:generic_moment_sym_scriptB}
    (\partial_t + p + c) A_{p,c} &= 
    -c (\B - 2c\beta) A_{p,c} \nonumber \\
    &\quad + [\B - (p+2c)\beta] (2c A_{p+1,c} - p A_{p-1,c+1}) \nonumber \\
    &\quad + c^2 \beta (2 A_{p+2,c-1} - A_{p+1,c-1}).
\end{align} A quantity of interest can be expressed in terms of the moments $ A_{p,c}$ and the parameters $\B$ and $\beta$. For example, the power emitted takes the form, 
\begin{equation}\label{eq:perm_symm_power}
    P = \beta (N A_{1,0} + N N_1 A_{0,1}) = \B A_{1,0} + \B^2 A_{0,1}/\beta -\B A_{0,1},
\end{equation} so to find $P$ to zeroth order, we need to find $A_{1,0}$ to zeroth order and $A_{0,1}$ to first order. Similarly, the coherence $G^{(2)}$ takes the form
\begin{align}\label{eq:perm_symm_G2}
    4G^{(2)} &= \beta^2 (2N N_1 A_{2,0} + 4N N_1 N_2 A_{1,1} + N N_1 N_2 N_3 A_{0,2}) \nonumber \\
    & = 2\B^2 A_{2,0} + 4\B^3 A_{1,1}/\beta + \B^4 A_{0,2}/\beta^2 + \mathcal{O}(\beta), 
\end{align} so to find $G^{(2)}$ to zeroth order, we need to find $A_{2,0}$ to zeroth order and $A_{1,1}$ to first order and $A_{0,2}$ to second order. The factor of $4$ on the left hand side is to account for both right and left propagation. For both $P$ and $G^{(2)}$, we note that we need $A_{p,c}$ to order $\beta^c$ to get $P$ or $G^{(2)}$ to order $\beta^0$.

Obtaining $A_{p,c}$ to order $\beta^c$ by a MF2 method can be shown to be exact. For example, in Sec.~\ref{sec:continuum_approximation_sym}, we employ the approximation $A_{1,1} \approx A_{0,1} A_{1,0}$. This forms a truncated system for the 3 variables $A_{0,1}$, $A_{1,0}$, and $A_{2,0}$, which are denoting the variables in Sec.~\ref{sec:continuum_approximation_sym}. The resulting solution can then be shown to be exact to order $\beta^c$ for a quantity $A_{p,c}$ as follows.

First, we note that the initial conditions propagate in order $\beta$ steps from $c = 0$ to higher $c$. This means that $A_{p,c}$ is of order $\beta^c$. Then, a solution ansatz exists in the form 
\begin{equation}\label{eq:ansatz_proof}
A_{p,c} = c e^{-pt}(A_{0,1})^c + \mathcal{O}(\beta^{c+1}),
\end{equation}
 which assumes a MF2 approximation to $\beta^{c}$ order, and $A_{0,1}$ has the same solution as the quantity $C(t)$ in \Eq{\ref{eq:C_xyt_sym}}.
\begin{equation}\label{eq:A0,1}
    A_{0,1}(t) = \frac{\beta}{\B}[\exp\{\B h(t)\}-1] + \mathcal{O}(\beta^2),
\end{equation} where $h(t)$ is from \Eq{\ref{eq:h}}. The ansatz $A_{p,c} = c e^{-pt}(A_{0,1})^c$ can be shown to be exact to order $\beta^c$ after a direct substitution in \Eq{\ref{eq:generic_moment_sym_scriptB}}. This provides a proof of the exactness of the MF2 in the thermodynamic limit considered in this work. Beyond the thermodynamic limit, i.e., for finite but large $N$ or for fixed but small $\beta$, going beyond MF2 is necessary to capture next order corrections as numerically shown in App.~\ref{app:beyond_thermodynamic_limit}.

\section{Beyond the thermodynamic limit}\label{app:beyond_thermodynamic_limit}
In this section, we summarize the order of MF$-n$ that is exact for various quantities of interest: the excitation, $e(t)$, the power, $P(t)$, second order coherence, $G^{(2)}(t,t)$, and the normalized second order coherence, $g^{(2)}(t,t)$. Additionally, beyond the thermodynamic limit, we find going to a higher order mean field method is necessary. For a quantity $F$ when starting from the excited state $\allexcited$, we have the expansion, $F = F_0 + \beta F_1 + \mathcal{O}(\beta^2)$. This is equivalent to a finite $N$ expansion, $F = F_0 + (1/N) \B F_1 + \mathcal{O}(1/N^2)$. Table~\ref{tab:minimum_mf_order} summarizes these results.


\begin{table}[t]
    \centering
    \caption{\justifying Minimum order $n$ of MF-$n$ required for the exactness of various $(1/N)$ expansion orders of the excitation $e(t)$, the power $P(t)$, second order coherence, $G^{(2)}(t,t)$, and the normalized second order coherence, $g^{(2)}(t,t)$. The asterisk (*) denotes unexplored quantities. The last column is a conjecture (denoted by `conj.') based on the low-order cases.}
    \label{tab:minimum_mf_order}
    \begin{tabular}{|c|c|c|c|c|c|}
    \hline
    \textbf{Quantity $\backslash$ Order} 
    & $1$ 
    & $\frac{1}{N}$ 
    & $\frac{1}{N^2}$ 
    & $\frac{1}{N^3}$ 
    & $\frac{1}{N^m} (conj.)$ \\ 
    \hline
    $e$ & 1 & 2 & 3 & * & $m+1$ \\
    \hline
    $P$ & 2 & 3 & 4 & * & $m+2$ \\
    \hline
    $G^{(2)}$ & 2 & 4 & 5 & * & $m+3$ ($m \ge 2$) \\
    \hline
    $g^{(2)}$ & 1 & 4 & 5 & * & $m+3$ ($m \ge 2$) \\
    \hline
    \end{tabular}
\end{table}

\begin{figure}[htbp]
    \centering
    \includegraphics[width=0.4\textwidth]{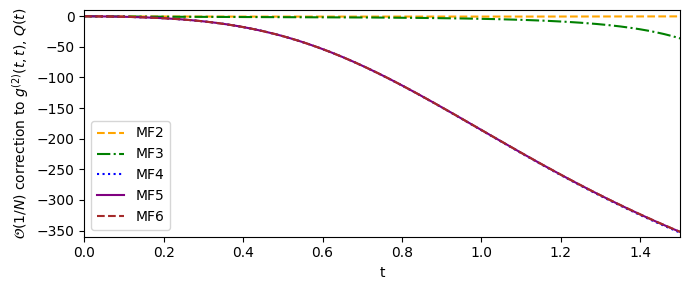}
    \caption{\justifying First order $(\mathcal{O}(1/N))$ correction, $Q(t)$, of the second order coherence $g^{(2)}(t,t)$ at $\B = N\beta = 10$ with $N = 2^{20}$ for a symmetric system using MF2 to MF6.}
    \label{fig:first_order_correction_g2}
\end{figure}

These results were obtained by a numerical simulation of the \emph{symmetric} system. For example, to verify that at least a fourth order is needed for the $(1/N)$ correction for $g^{(2)}$, we simulate a large number $N = 2^{20}$ with a fixed $\B = N\beta = 10$ using MF2 to MF6 methods. We then track $Q(t) = (g^{(2)}(t,t) - 2)/\beta$. This is shown in Fig.~\ref{fig:first_order_correction_g2}. For MF4 and beyond, the quantities exactly agree together. This is evidence for the exactness of MF4 and beyond for $Q(t)$, while MF2 and MF3 fail to predict it. We note that this behavior is similar to the finite $N = 300$ analysis shown in the appendix of the companion paper \cite{tohfachiral}. All the other quantities explored in Tab.~\ref{tab:minimum_mf_order} were similarly checked by simulations but not are not shown.

General features of the various corrections were observed. First, the corrections alternate in sign. For example, $P_1$ is negative, while $P_2$ is positive. Second, a higher order correction is only significant for later times. Third, the corrections get larger in magnitude with increasing $\B$.




\bibliography{main}

\end{document}